\documentclass[manuscript]{acmart}

\AtBeginDocument{%
  }

\setcopyright{acmlicensed}
\copyrightyear{2018}
\acmYear{2018}
\acmDOI{XXXXXXX.XXXXXXX}

\acmConference[Conference acronym 'XX]{Make sure to enter the correct
  conference title from your rights confirmation email}{June 03--05,
  2018}{Woodstock, NY}
\acmISBN{978-1-4503-XXXX-X/18/06}

\begin{document}

\title{Generative Ghosts: Anticipating Benefits and Risks of AI Afterlives}

\author{Meredith Ringel Morris}
\email{merrie@google.com}
\orcid{0000-0003-1436-9223}
\affiliation{%
  \institution{Google DeepMind}
  \city{Seattle}
  \state{Washington}
  \country{USA}
}

\author{Jed R. Brubaker}
\email{jed.brubaker@colorado.edu}
\orcid{0000-0003-4826-8324}
\affiliation{%
  \institution{University of Colorado Boulder}
  \city{Boulder}
  \state{Colorado}
  \country{USA}
}

\renewcommand{\shortauthors}{Morris and Brubaker}

\begin{abstract}
  As AI systems quickly improve in both breadth and depth of performance, 
 they lend themselves to creating increasingly powerful and realistic agents, 
 including the possibility of agents modeled on specific people. We anticipate that within our lifetimes it may become common practice for people to create custom AI agents to interact with loved ones and/or the broader world after death; indeed, the past year has seen a boom in startups purporting to offer such services. We call these \textit{generative ghosts}, since such agents will be capable of generating novel content rather than merely parroting content produced by their creator while living. In this paper, we reflect on the history of technologies for AI afterlives, including current early attempts by individual enthusiasts and startup companies to create generative ghosts. We then introduce a novel design space detailing potential implementations of generative ghosts, and use this analytic framework to ground discussion of the practical and ethical implications of various approaches to designing generative ghosts, including potential positive and negative impacts on individuals and society. Based on these considerations, we lay out a research agenda for the AI and HCI research communities to better understand the risk/benefit landscape of this novel technology so as to ultimately empower people who wish to create and interact with AI afterlives to do so in a beneficial manner.    
\end{abstract}




\maketitle

\section{Introduction}
\label{intro}

The past few years have brought incredible growth in the capabilities of generative AI models, particularly large language models (LLMs) such as GPT-4 \cite{openai2023gpt4}, Palm 2 \cite{palm2}, and Llama 2 \cite{llama2}, though there has also been incredible progress in generative AI for the production of images \cite{dalle2}, video \cite{singer2022makeavideo}, and audio \cite{audiolm}, as well as a new generation of multimodal models \cite{dawnoflmms, gemini} that combine functionality across several media categories. These models, in turn, have given rise to new types of \textit{generative agents} \cite{generativeAgents}, simulacra that can produce believable human behaviors, including capabilities such as memory and planning. While still in their infancy, generative agents and related technologies are likely to increase in fidelity and popularity as underlying model capabilities improve, compute costs drop, and the required technical expertise decreases. For instance, in November 2023 OpenAI released GPTs \cite{openAIBlog}
, a no-code interface for people to develop agentic AIs.   
As AI models increase the set of human capabilities they can faithfully reproduce \cite{morris2023levels, bubeck2023sparks}, societal change is inevitable. For instance, experts anticipate that powerful AI systems may profoundly change disparate areas of society such as the labor market \cite{eloundou2023gpts}, the education system \cite{llmsEd}, the pursuit of scientific knowledge \cite{morris2023scientists}, and criminal activities \cite{ferrara2023genai}. In this paper, we discuss how advances in AI might change personal and cultural practices around death and dying as well.

We introduce the concept of \textit{generative ghosts}\footnote{We use the term "ghosts" to encapsulate both the memorializing practices we imagine these agents may fulfill and the unintentional persistent presences we anticipate such AI agents will bring.}, AI agents that represent a deceased person, and discuss why we anticipate such representations may become increasingly common. We share examples from recent media reports of individuals and startups experimenting with the creation of generative ghosts, a trend that seems to be accelerating, particularly in East Asia. We introduce a design space to describe possible instantiations of generative ghosts and use this analytic framework to explore how design choices might lead to practical and ethical concerns as well as potentially beneficial outcomes.  

Our contributions include: (1) identifying and characterizing an emerging phenomenon of creating ``generative ghosts'' to represent the deceased, (2) introducing a design space of dimensions and analysis of potential benefits and harms that can be used to support future empirical research and motivate fieldwork. By characterizing this emerging trend, highlighting potential risks, and creating a framework for future investigation, we aim to ensure that future technical and sociotechnical systems will maximize the potential benefits of ``AI afterlives'' while minimizing potential risks.

\section{Related Work}
\label{related}

We discuss the rich literature on how technologies have changed practices around death and dying and initial forays into AI afterlives by individuals and start-up ventures. 

\subsection{Post-Mortem Technology}



Throughout history, people have turned to technology to remember, memorialize, and even interact with the dead. Gravestones and other burial markers can be traced nearly back to 3000 B.C.E. \cite{taylorDeathAfterlifeAncient2001}. Obituaries in the U.S., while dating back to the 16th century, became more common during the 19th century in part due to the U.S. Civil War \cite{humeObituariesAmericanCulture2000} -- an event that also brought embalming into favor. Even the mediums of the Spiritualism movement in the late 19th and early 20th century turned to telegraphs, radio-wave detectors, and later wireless radio in their attempts to detect the presence of and communicate with the dead \cite{nationalscienceandmediamuseumTelecommunicationsOccult2022}.

During the earliest days of the World Wide Web, when people would create personal Home Pages describing their lives and family, it was routine for people to dedicate a page to the memory of a family member, often a deceased parent or family pet \cite{brubakerDeathIdentitySocial2015}. Online graveyards, websites specifically dedicated to the memorialization of the dead, soon followed \cite{robertsPerpetualCareCyberspace2000, robertsLivingDeadCommunity2004}. The \textit{Virtual Memorial Garden} is the earliest documented example \cite{robertsPerpetualCareCyberspace2000}. Created in 1996, it featured a collection of brief obituaries authored by loved ones, capturing their grief, the circumstances around the death, and loved ones' guilt.

Even as people adopted digital technology to memorialize the dead, scholars in human-computer interaction and social computing began to note how mortality is overlooked by technology design and have argued for increased attention to end-of-life and mortality \cite{massimiDyingDeathMortality2009,massimiHCIEndLife2010,odom2010, bellNoMoreSMS2006}, often termed ``thanatosensitive design'' \cite{massimiDyingDeathMortality2009}. Scholars have studied the intersection of technology and mortality across diverse contexts, including digital heirlooms (e.g., \cite{odom_technology_2012,beuthelExploringBodilyHeirlooms2022}), communal rituals (e.g., \cite{hakkila_designing_2019,uriu_floral_2021}), online memorials (e.g., \cite{gulotta_legacy_2014,moncur_emergent_2014,massimi_exploring_2013}), digital legacy (e.g., \cite{gulotta2016engaging,gulotta_legacy_2014,gulotta2017digital,Doyle2023-digital-legacy-SLR}), and family archives (e.g., \cite{kaye_have_2006}). 
Yet, as Doyle and Brubaker summarize \cite{Doyle2024-ADP}, end-of-life scenarios also present new privacy challenges (e.g., \cite{holt_personal_2021,locasto_security_2011}), challenges that result from shifting motivations at different life stages \cite{chen_what_2021,thomas_older_2014}, and challenges due to differences between user expectations and platform functionality \cite{gach_experiences_2020}.

In HCI, extensive attention has been given to \textit{digital legacies}, the collection of materials that carry values and meaning that are passed down and/or otherwise continue to represent the deceased after their death. A recent literature review of the field by Doyle and Brubaker \cite{Doyle2023-digital-legacy-SLR} identified four foci of this scholarship: digital identity (i.e., how legacies can continue to represent the deceased in intentional and unintentional ways), engagement with digital legacies (i.e., studies on use and user perceptions), putting to rest (i.e., concerned with issues around preservation, disposition, and disposal), and the integration of technology into existing legacy practices \cite{Doyle2023-digital-legacy-SLR}.

In most cases, it is the digital traces people leave behind that give rise to legacy crafting and memorialization opportunities. The sources of legacy content vary -- from burner accounts \cite{gulotta2016engaging} to personal archives \cite{she_living_2021, kaye_have_2006} -- but social media content (especially memorialized profiles) has received the most scholarly attention
(e.g., \cite{brubaker_stewarding_2014,gach_experiences_2020,gach_getting_2021,brubaker_legacy_2016,brubaker2011we33,brubaker2019orienting,getty2011said,mori2012design,sofkaAdolescentsTechnologyInternet2009,sofkaGriefAdolescentsSocial2017,arnoldDeathDigitalMedia2017,nansenDeathTwitterUnderstanding2019}). 

The persistence of identities has benefits and challenges: Digital legacies can facilitate healthy grieving \cite{she_living_2021,brubakerGraveFacebookSite2013,brubakerGriefStrickenCrowdLanguage2012} and maintain connections to the deceased \cite{brubaker2011we33,getty2011said,gulotta_legacy_2014}. However, receiving a large and uncurated set of content can be overwhelming for loved ones \cite{holt_personal_2021}, and may provide (for better or worse) an uncensored version of the deceased \cite{gulotta2013digital}. Loved ones may also find themselves overwhelmed as they become the contact point for the deceased's various online networks \cite{brubaker_stewarding_2014, brubakerGraveFacebookSite2013}.

Across many studies, participants describe uncertainty around the use, care, and understanding of digital legacies. Ambiguities often arise when data created for one purpose -- e.g., personal archives \cite{kaye_have_2006,holt_personal_2021} or interpersonal communication \cite{moncur_emergent_2014,thomas_older_2014,gach_getting_2021,gach_experiences_2020,gulotta_legacy_2014} -- are repurposed post-mortem. In contrast to traditional legacies where there are familiar norms and practices, Pfister \cite{pfister2017will} argues that common norms around digital legacy have not yet emerged, an issue that AI will certainly exacerbate.

\subsection{AI Afterlives}


A handful of tech-savvy individuals have attempted to create interactive memorializations of loved ones by ingesting their digital (or digitized) content such as emails, journals, videos, photographs, or other autobiographical media. In her graphic novel \textit{Artificial: A Love Story}, Amy Kurzweil \cite{artificial} documents how her father, futurist Ray Kurzweil, created a chatbot to embody the memory of his deceased father, Fred Kurzweil, in a form he dubs ``Fredbot'' \cite{kurzweilNews}. 
Fredbot interactively responds to questions from his descendants, but only by sharing exact quotes from the materials Fred left behind such as letters digitized by his family. In another well-publicized incident, the engineer Eugenia Kuyda created an app to preserve the memory of her best friend, Roman, who died unexpectedly in an accident, training a neural network on text messages her friend had sent her to create a bot named, suitably, ``Roman.'' Unlike Fredbot, which was available only to immediate family, the Roman bot was made available on social media and app stores 
for public interaction, resulting in mixed reactions from friends and family of the deceased \cite{romanNews}. 
While Fredbot and Roman bot represented private citizens, AI has also been used to re-create public figures in various ways. For instance, in November 2023, the surviving members of The Beatles released a new song, ``Now and Then,'' \cite{beatles} using AI to enable the deceased John Lennon to sing along with his living bandmates \cite{beatlesNews}. The musician Laurie Anderson collaborated with the University of Adelaide’s Australian Institute for Machine Learning to create a chatbot based on her longtime partner, deceased musician Lou Reed, which, in addition to conversing, generates new music lyrics in Reed's style \cite{reedNews}. In early 2024, gun-control activists in the United States used AI to re-create the voices of victims of gun violence, to create a visceral and poignant plea to U.S. legislators \cite{gunNews}.

Meanwhile, there are a handful of start-up companies that aim to give people the ability to create their own AI afterlives while they are still alive. Re;memory\footnote{rememory.deepbrain.io/en} is an offering from DeepBrain AI that professes to create an interactive virtual representation after a seven-hour filming and interview session. They specifically advertise this service as a way that one can proactively create a rich memory that friends and family can engage with after one dies. HereAfter\footnote{https://www.hereafter.ai} provides an app that interviews a user with the goal of proactively creating a posthumous digital representation. Friends and relatives can interact with a chatbot-based representation of their loved one who can provide photos and voice recordings of memories and life events. Early attempts at generative ghost technologies seem to already be entering the mainstream in East Asia, where "communicating" (via non-technological means) with deceased ancestors 
 is already a cultural norm for many \cite{chinaDeepFakes}.  Companies offering "digital immortality" are increasingly popular in China, serving thousands of customers \cite{chinaGhosts, nprChina, chinaDeepFakes}. In South Korea, 19 million people have streamed an emotional video of a bereaved mother, Jang Ji-Sun, interacting with a virtual reality representation of her deceased young daughter, Na-yeon, custom-created for her by a media company \cite{koreanDocumentary}.

In contrast with Re;memory and HereAfter, which aim to empower users to preserve their memories before their deaths, some start-ups purport to use AI to provide an experience more akin to resurrection or reanimation (i.e., without the direct involvement of the representee, and often after their death). For instance, Character.AI\footnote{https://beta.character.ai/} uses LLMs to simulate chatting with deceased public figures for entertainment or education (e.g., a bot representing William Shakespeare), and the Musée D'Orsay in Paris recently contracted with a start-up company to develop a chatbot meant to represent the artist Vincent van Gogh by ingesting his writings \cite{vangogh}. Beyond public figures, Project December\footnote{projectdecember.net/} is a venture that allows end-users to upload content from a deceased loved one \cite{decemberNews} in order to ``simulate the dead'' via a customized chatbot. The Augmented Eternity project at MIT is an academic endeavor that aims to allow people to create digital representations of themselves with the express purpose of being able to agentically represent them after death to members of their social network \cite{augmentedEternity}.

While the examples shared above demonstrate the diversity of emerging approaches, a key trend in post-mortem AI is around griefbots. \textit{Griefbots} \cite{griefbot1, griefbot2}, sometimes also called  \textit{deathbots} \cite{deathbots} or \textit{deadbots} \cite{hollanekGriefbotsDeadbotsPostmortem2024}, are typically described as chatbots (predating modern large language model technologies) designed to allow the bereaved to ``converse'' with the deceased, typically created posthumously by a third party who leverages general-purpose text the deceased created during their life. The aforementioned ``Fredbot'' and ``Roman'' would be considered examples of griefbots. 

Because of their third-party origins, griefbots raise controversial ethical issues regarding privacy and consent \cite{Grandinetti_DeAtley_Bruinsma_2020} which must be considered from multiple perspectives.
For example, Hollanek and Nowaczyk‑Basińska \cite{hollanekGriefbotsDeadbotsPostmortem2024} emphasize the importance of considering the complex relationships among three classes of stakeholders for such bots: data donors, data recipients, and service interactants. The complexity they raise aligns with our own thinking, both in our design framework (see first- vs. third-party generative ghosts) and in our discussion of risks and benefits. However, we use the terms \textit{deceased} and \textit{bereaved} to emphasize interpersonal and relational aspects of clones, while also considering the impacts on society at large. 

Generative ghosts can be considered an extension of the concept of a griefbot, but not all griefbots would be considered generative ghosts. Generative ghosts go beyond the traditional definition of griefbots in several ways. Generative ghosts always include the ability to generate novel content in-character (whereas many griefbots only regurgitate content from a training corpus) and potentially evolve over time. For instance, a user should be able to ask a generative ghost questions about current events and obtain responses that would be ``in character'' for the deceased even though these events occurred after the end of the representee's life. Additionally, generative ghosts extend the notion of griefbots by possessing agentic \cite{generativeAgents, openAIagentic} capabilities such as the ability to participate in the economy or perform other complex tasks with limited oversight (in AI parlance, generative ghosts can engage in \textit{tool use} \cite{bubeck2023sparks}). Notably, our framework for generative ghosts envisions the possibility of first-party (rather than only third-party) creation as part of an end-of-life planning process. Another key expansion to the notion of griefbots is that generative ghosts might exist \textit{pre-mortem} as a generative clone (an AI agent representing a living person), which then transitions into a generative ghost upon the representee's death.  Finally, in contrast to griefbots, generative ghosts may be created for reasons other than to support the bereaved in their grieving process. The following section introduces more details about the design space of generative ghosts. 

\section{Generative Ghosts: A Design Space}
\label{ghosts}

We introduce the notion of a \textit{generative ghost}, an agentic, AI-powered representation of a deceased individual. Drawing on methods from design research (e.g., \cite{designResearchCMU}), we used an analytical approach to generate a design space. Our analysis started with a growing set of individual attempts to create generative ghosts as reported in the news media and descriptions of start-up companies purporting to offer related services (see Section 2). Drawing on our respective expertise as scholars in technical human-AI interaction (first author) and in thanatosensitive design (second author) we considered those systems and services in the context of related work in our respective areas, speculating on possible and likely futures. As such, our proposed framework is the byproduct of analytic and conceptual explorations drawing on related work and systems in the wild to date. We share this framework with the hopes that it will guide future work while also knowing that future design and empirical work will likely result in updates to our proposed framework.

Generative ghosts differ from other technology-mediated representations of the deceased (such as memorialized social media accounts) in that they provide more than static or interactive access to content about the deceased produced during their lifetime (e.g., text, imagery, audio, video, or other media). Rather, they additionally are capable of generating novel content and interactions based upon the deceased's data, potentially mediated by additional metadata such as personality questionnaires \cite{gss}, prompts \cite{cantPrompt}, constitutions \cite{bai2022constitutional}, or other rules set forth by the generative ghost's creator. In addition to producing novel content, generative ghosts may also possess \textit{agentic} \cite{generativeAgents, openAIagentic} capabilities (i.e., the ability to autonomously execute complex sequences of actions with side effects in either the digital or physical realms).

Generative ghosts can be considered a special case of a broader class of agentic generative AI systems meant to represent a specific individual, which we will call \textit{generative clones}. Generative clones are agents created for or by a living person to mimic their persona in order to execute actions and interactions on their behalf during their lifetimes. For example, a busy professional might create a generative clone of herself to respond to low-priority emails or phone calls in a manner that mimics her writing style, voice, and other characteristics such that her correspondents would think she had personally crafted the replies. Using this terminology, a generative ghost can be defined as a generative clone representing a deceased person.

We expand on our definition of generative ghosts by discussing the design space of potential instantiations of this concept. 
While all generative ghosts share the capability to retrieve pre-existing multimodal content and generative novel multimodal content representing a deceased individual (including through the execution of agentic interactions), we argue that several key dimensions may impact the capabilities, perceptions, and societal impacts of such systems, including \textit{provenance}, \textit{deployment timeline}, \textit{anthropomorphism paradigm}, \textit{multiplicity}, \textit{cutoff date}, \textit{embodiment}, and \textit{representee type}. As we detail in the Discussion, interrelationships between these dimensions are common. However, we present them separately here for the sake of clarity.


\subsection{Provenance: First-Party vs. Third-Party Ghosts}

The \textit{provenance} of a generative ghost refers to who created it (or authorized its creation). A \textit{first-party} generative ghost is one created by the individual being represented (i.e., perhaps via end-of-life planning or legacy crafting practices). First-party ghosts may also be created by the individual, but not explicitly for legacy purposes (akin to how social media profiles may become memorials after someone dies \cite{Doyle2023-digital-legacy-SLR}). In contrast, a \textit{third-party} generative ghost is not created by the representee. Third-party generative ghosts might be created by parties with a direct personal connection to the deceased (e.g., family, friends), by parties with a fiscal connection to the deceased (e.g., employers, estates), or by unconnected parties. Additionally, third-party ghosts may be authorized or unauthorized by the representee. 
Authorized third-party generative ghosts, for example, might be created with consent (e.g., via the deceased's will). Meanwhile, unauthorized ghosts created by unconnected third parties are most likely in the case of public figures such as historical figures or contemporary celebrities; indeed, we already see early attempts at creating third-party generative ghosts of historical figures for entertainment and educational purposes by companies such as character.ai \cite{charNews} and nonprofits such as Khan Academy \cite{khanNews}.

\subsection{Deployment Timeline: Pre- vs. Post-Mortem}

Closely related to provenance, the \textit{deployment timeline} of a generative ghost indicates whether the system's initial deployment was \textit{pre-mortem} or \textit{post-mortem}. While some generative ghosts will be deployed post-mortem with the explicit purpose of memorializing the dead, many generative ghosts will likely start off as agents that serve some pre-mortem functionality (e.g., a generative clone responding to personal emails or accomplishing other tasks on behalf of the representee, in character as if the representee had executed the task themself) only to transition to ghosts after the death of their representee. Similar to digital legacy planning \cite{Doyle2024-ADP, Doyle2023-digital-legacy-SLR}, pre-mortem deployments might be desirable as they would allow an individual the opportunity to tune the behavior and/or capabilities of a first-party generative ghost to their liking, perhaps as part of an end-of-life planning process. 
We also expect that generative clones would benefit from being designed with mortality in mind, in line with calls for thanatosensitivity in other technology \cite{massimiDyingDeathMortality2009}. The design of clones should likely include modifications to their behavior and/or capabilities once they become ghosts.\footnote{Note that the ethical implications of \textit{pre-mortem} generative clones (also called ``digital twins,'' ``AI clones,'' or ``mimetic models'') have been discussed in related works such as \cite{mimeticModels, aicloneRisks}; our contribution focuses on identifying potential risks, benefits, and a research agenda around \textit{post-mortem} clones, i.e., generative ghosts.}


\subsection{Anthropomorphism Paradigm: Reincarnations vs. Representations}

The \textit{anthropomorphism paradigm} of a generative ghost refers to a subtle, but potentially impactful, interface metaphor choice: whether a generative ghost presents itself as a \textit{reincarnation} of the deceased individual or as a \textit{representation} of that individual. Generative ghosts employing a reincarnation metaphor would engage with end-users as if they \textit{were} the deceased individual. In contrast, generative ghosts employing a representation metaphor would engage with end-users as if they are a representation of the deceased, but are not the deceased themselves. We see similar distinctions in the context of social media where HCI literature and popular press have documented scenarios where platforms animate users even after their deaths (e.g., through algorithmically generated notifications \cite{Jiang2018-ripgenre}) and the need to be clear about whether content is the result of a user's actions or the platform's \cite{brubakerGraveFacebookSite2013}. In the context of generative ghosts, some design choices that might impact perceptions of reincarnation vs. representation include whether a generative ghost uses first-person pronouns, whether it uses the present or past tense when discussing its representee, whether it uses the name of its representee or a different nomenclature (e.g., the use of ``Fredbot'' in \cite{artificial}), and whether it is allowed to make statements that assert it is alive, possesses a soul, etc.


\subsection{Multiplicity: Single Ghost vs. Multiple Ghosts}

The \textit{multiplicity} design dimension refers to whether an individual is represented by a \textit{single} generative ghost or by \textit{multiple} distinct ghosts. Multiple ghosts might arise intentionally (i.e., a single creator develops a set of generative ghosts with different behaviors, capabilities, and/or audiences in mind). Multiple ghosts might enable people to customize different ghosts for different audiences in order to avoid forms of context collapse \cite{davisContextCollapseTheorizing2014,littKnockKnockWho2012a,marwickTweetHonestlyTweet2011}. Multiple ghosts might also arise unintentionally due to multiple third parties creating generative ghosts for a single individual, or perhaps due to post-mortem identity theft or other novel cybercrimes. 

\subsection{Cutoff Date: Static vs. Evolving}

The \textit{cutoff date} of a generative ghost describes whether the ghost remains \textit{static} (i.e., not developing novel interests, skills, or other characteristics after its representee's death) or whether it is \textit{evolving}. Evolving ghosts' characteristics might change over time, such that they may eventually diverge in substantial ways from the deceased individual. For example, consider a hypothetical situation in which a parent creates a third-party generative ghost to represent a deceased child. A static cutoff-date design would result in a representation that perpetually interacted in a style faithful to the appearance, diction, maturity, etc. of a young child, whereas an evolving representation might ``age'' (either in real-time or per some other scheme). 

While the previous example speaks to whether a ghost is designed to evolve by simulating aging, a ghost could also evolve (albeit in less predictable ways) if new information about the deceased is added to the model (e.g., in the case that additional data about the deceased was found). Moreover, the informational context beyond data about the representee (e.g., from world news to the personal information of those who might interact with the ghost) can be static or be allowed to evolve. One can imagine scenarios in which a ghost might be designed to incorporate (or not) new information, including personal information (such as a wedding or birth of a child) or world events (such as a future election).\footnote{In 2013, a now-defunct service called LivesOn aimed to approximate a rudimentary version of this concept. The service promised to animate a deceased user's Twitter profile with tweets written in their likeness, commenting on the latest news headlines \cite{LivesOn}.}

\subsection{Embodiment: Physical vs. Virtual Ghosts}

The \textit{embodiment} dimension refers to whether a generative ghost has a \textit{physical} embodiment. Such embodiments might be physical in a literal sense (e.g., robotics) or might be embodiments in rich digital media (e.g., avatars in mixed reality environments). In contrast, purely \textit{virtual} generative ghosts would lack embodiment (e.g., perhaps available only as a chatbot). In addition to technical and/or cost constraints, there may be additional reasons to opt for lower-fidelity (i.e., virtual-only) embodiments, including potential ethical or psychological concerns related to physical embodiments. 

\subsection{Representee: Human vs. Non-Human Legacies}

The \textit{representee type} dimension acknowledges that, in addition to representing deceased \textit{humans}, people may wish to create ghosts representing \textit{non-humans}, such as beloved family pets or service animals. While this paper is primarily concerned with generative ghosts representing deceased people, we also recognize the need for reflection on technology to support interactive remembrances of non-humans.  

\section{
Benefits and Risks of Generative Ghosts}
\label{benfitsAndRisks}

Having introduced the concept of generative ghosts and the dimensions associated with this design space, we now discuss the potential benefits of thoughtfully and safely implemented generative ghosts, followed by a discussion of potential risks associated with various implementations.  

\subsection{Generative Ghosts: Potential Benefits}
\label{benefits}

We first consider likely motivations for creating generative ghosts. With appropriate attention to technical concerns (e.g., interaction design, AI safety, etc.) and sociotechnical concerns (e.g., ethics, cultural concerns, legal ramifications, etc.), one can envision a positive future in which generative ghosts offer a host of benefits both to the representee, the bereaved, and society more broadly.

\subsubsection{Potential Benefits for Representees}

Individuals may choose to create first-party generative ghosts as part of their end-of-life planning for several reasons. On a personal level, people may feel comforted by or even feel hopeful about the idea of a ``digital afterlife.'' Prior work on digital legacy has documented people's motivations to capture their life stories \cite{thomas_older_2014}, major life events \cite{jamison-powell_ps_2016}, and pass on values to future generations \cite{gulotta2017digital}. 
Posthumous, personalized AI representations may likewise be seen as a way for people to ensure that they are remembered, whether by family, friends, or perhaps the world at large. 

The likely nature of generative ghost architectures makes such remembrances higher-fidelity, more interactive, and more personalized 
than today's more common digital memorials, such as memorialized social media accounts. As such, generative ghosts may help give individuals a sense of agency over their posthumous future, such as by presenting an ability to comfort loved ones after they have passed, to offer advice to loved ones at future events (e.g., convey well wishes and advice upon the marriage of a great-grandchild), or to preserve their personal, religious, and/or cultural heritage by sharing knowledge and wisdom specific to their background and time in history to future generations. 

It is even possible that generative ghosts may offer legal or economic benefits to representees. For example, one can imagine that creating a generative ghost might replace or complement life insurance policies for some individuals if their AI embodiments can participate in our economic system and earn an income for their descendants (e.g., an author whose generative ghost continues to produce novel works in their style). Depending on how our legal system evolves, generative ghosts may also provide utility in ensuring the proper execution of other aspects of end-of-life planning, such as arbitrating disputes over interpretations of a will. 

\subsubsection{Potential Benefits for the Bereaved}
Interacting with a generative ghost may provide emotional support to close friends and family of the deceased. Prior research has considered the impact of online memorials on the bereaved (e.g. \cite{walterDoesInternetChange2012}), often responding to concerns that such spaces might prolong grief (see Section \ref{mentalHealthRisks}). Yet scholars cite the benefits of online memorials that allow loved ones to maintain important ``continuing bonds'' \cite{klassContinuingBondsNew1996} with the deceased (e.g., \cite{getty2011said, brubaker2011we33, hjorth2021place}; c.f., \cite{degrootModelTranscorporealCommunication2018}), often in a space where those grieving can gather \cite{carroll2010, brubakerGraveFacebookSite2013}. 

As with griefbots more generally, discussions with a generative ghost might provide comfort or closure and support a continued feeling of closeness. Generative ghosts may support positive mental health, particularly if they enable the bereaved to maintain connections with the deceased while ``making sense'' of the loss (termed ``accommodation'' \cite{neimeyerMeaningMakingArt2014}).
Family members may find it comforting not only to interact with the generative ghost in the immediate post-mortem period, but to know that they might be able to interact with their deceased loved one at upcoming future life events such as weddings, the birth of a child, or other milestones during which they may wish to share updates with the generative ghost and/or hear advice, congratulations, or other event-appropriate commentary from their deceased loved one. 

One-sided conversations with deceased loved ones, such as speaking graveside or writing to the deceased on a memorial, are common and well-recognized in the literature \cite{degrootModelTranscorporealCommunication2018, brubaker2011we33, brubakerGraveFacebookSite2013} and are generally considered a part of healthy grieving across many grief paradigms (e.g.,  \cite{klassContinuingBondsNew1996}). As summarized by DeGroot \cite{degrootModelTranscorporealCommunication2018}, 
these conversations are cathartic and help the living confront their loss: ``People essentially struggle to develop an identity where the deceased is now part of the past self and not the present self... and communicating with the deceased appears to ease the transition'' \cite{degrootModelTranscorporealCommunication2018}. However, it is worth noting that the dead do not typically respond. Rather, people imagine how the deceased would reply \cite{degrootModelTranscorporealCommunication2018}. As such, it is unclear how a generative ghost that can respond might impact bereavement.

Finally, generative ghosts may also embody practical knowledge related to end-of-life planning in a method that is easily discoverable to and digestible by surviving relatives, such as helping convey information to support funeral or memorial planning or supporting discovery of items associated with transitions of ownership of financial accounts or other property such as passwords or locations of key items. Generative ghosts might also support family members by providing advice on procedures that they had been responsible for in life (e.g., teaching a surviving spouse how to cook a favorite dish or repair the kitchen faucet). In some cases, income provided by generative ghosts' participation in the economy might support family members.  


\subsubsection{Potential Benefits for Society}

The interactive modality of generative ghosts may provide benefits, particularly in the cases of culture, history, and heritage. As mentioned above, organizations such as character.ai and Khan Academy are already considering the benefits for entertainment and education. 

Generative ghosts may also be beneficial for smaller cultural groups. They may be one way to preserve the collective wisdom of elders, as well as cultural heritage including the knowledge of dying languages, religions, or other small-membership cultural practices that are at risk of 
being forgotten as their members age. For instance, generative ghosts may be one way to preserve historical knowledge about events such as the Holocaust before the few remaining elderly survivors with firsthand experience pass away \cite{zuckerAIArchivesHow2023,colavizzaArchivesAIOverview2021}.

Indeed, the creation of generative ghosts may enrich the practice of disciplines such as museum curation, historical scholarship, anthropology, and other related humanities. These disciplines may benefit, not only from the static knowledge encoded in generative ghosts, but from the ability to interactively query such representations to understand perspectives and events in new ways. For example, devout adherents to a specific religious tradition might be able to converse with generative ghosts of deceased elders or scholars of their faith about changing aspects of society (e.g., perhaps an Orthodox Jew might engage with generative ghosts of Talmudic scholars to debate the pros and cons of whether kashrut allows consumption of synthetic, lab-grown pork). Generative ghosts might also represent archetypes or amalgamations rather than specific individuals, particularly when developed from historical records (e.g., ``citizen of Pompeii,'' ``typical resident of Colonial Williamsburg,'' etc.).

\subsection{Generative Ghosts: Potential Risks}
\label{risks}
While thoughtful implementations of generative ghosts have potential upsides for the representee, their loved ones, and society more broadly, there are also risks. By proactively anticipating these risks, we may be able to design technical and sociotechnical systems that avoid or mitigate them. Likewise, we might be able to identify applications for which the risks are too high, regardless of thoughtful design. Of course, as with any powerful new technology, it is impossible to fully anticipate apriori the many potential sociotechnial side-effects of its introduction. The emergence of ``AI afterlives'' may reshape society in complex ways beyond our current imagining (i.e., spurring the advent of new religious movements). Here, we discuss four broad categories of risk associated with generative ghosts: mental health risks, reputational risks, security risks, and socio-cultural risks.

\subsubsection{Mental Health Risks}
\label{mentalHealthRisks}
Earlier we discussed ways in which generative ghosts might offer mental health benefits to mourners. However, ghosts may also pose \textit{mental health risks} for the bereaved, some of which we outline in this section. It is important to note, however, that what is considered a mental health risk is context-dependent as grief paradigms vary extensively across cultures. For example, common Chinese rituals and practices related to ancestors --- practices that appear to be shaping the acceptability of generative ghosts already --- might be interpreted as problematic elsewhere. Similarly, the Māori of New Zealand see ancestors as kaitiaki (guardians) who offer spiritual support and ensure the well-being of their descendants \cite{schwass2005last}. As such, behavior considered risky in a Western context may be seen as normal or even beneficial in another.
The mental health risks we outline in this section are predominantly drawn from Western perspectives as they reflect the concerns most frequently raised in the literature related to technology and mortality. However, even as we enumerate potential risks, we would encourage readers to consider cultural factors that shape what is considered a risk vs. a benefit.

Western scholars in thanatology and clinical psychology have long distinguished between \textit{adaptive coping} -- strategies that integrate the loss into an individual's life, fostering emotional adjustment, well-being, and personal growth over time -- and \textit{maladaptive coping} -- behavior that may obstruct healthy grieving, prolonging distress and resulting in depression, anxiety, and complicated grief (also known clinically as Prolonged Grief Disorder)\cite{stroebe2013complicated, shear2015complicated}. 
Meanwhile, studies of prior generations of AI technologies (and even of non-AI computing systems \cite{nassMedia}) have found that people tend to anthropomorphize such systems and even develop attachments \cite{assistantEthics}. As such, ghosts may include risks related to delayed accommodation, complicated grief, information overload, anthropomorphism, deification, and second deaths, which we detail below. We anticipate that without careful design that generative ghosts might magnify risks due to the increasing fidelity of generative AI technologies and the modeling of the agent on a specific individual. 

Interacting with a generative ghost may impact the bereaved's ability to assimilate and accommodate \cite{accom, neimeyerMeaningMakingArt2014} their loss 
and integrate the reality of their loved one's death \cite{klassContinuingBondsNew1996}.
\textit{Delayed accommodation} may negatively impact the bereaved's mental health and quality of life by supporting loss-oriented experiences (e.g., active grieving, reminiscing while looking at old photos, etc.) at the expense of restorative-oriented experiences (e.g., the development of new relationships and life patterns)  \cite{stroebeDualProcessModel1999}. Stroebe and Schut argue that both loss- and restorative-oriented experiences are necessary, but that having agency around when and how one oscillates between the two is important -- agency that could be impeded by the interactive design of generative ghosts (e.g., if a generative ghost were designed to use ``push notifications'' that might pull a mourner into an interaction rather than letting the mourner decide when and how they want to engage in remembrance and other grief-related activities).  

A related concern is the possibility of \textit{complicated grief}, i.e., developing an overly strong reliance on generative ghosts as indicated by spending a disproportionate amount of time engaging with representations of the deceased rather than interacting with the living.
Already, there are indications that some individuals find AI-powered companions, such as those from Replika (replika.ai), to be highly compelling \cite{replikaNews}
; generative ghosts' basis in real, beloved individuals rather than in fictional personas or historical figures may amplify addictive risks of interacting with AI agents \cite{addictiveIntelligence}. 

People may also overly rely on a generative ghost's advice for momentous or even mundane life decisions. While soliciting input from ghosts might be positive or harmless in some circumstances, one can imagine how the use of ghosts could result in \textit{information overload}, lead to choice paralysis, or create tensions over how to weigh competing recommendations from a generative ghost vs. living friends and relatives. 

\textit{Anthropomorphism} of generative ghosts is a potential risk for mourners, if they become convinced that a generative ghost truly is the deceased (rather than a computer program representing them). Anthropomorphism might increase the likelihood of other mental health risks such as complicated grief or delayed accommodation \cite{shear2015complicated}. Some design choices we discussed earlier may increase anthropomorphism risk, particularly the use of reincarnation metaphors, evolving representations, and physical embodiments. 

\textit{Deification} is a more extreme version of the anthropomorphism risk, in which a mourner might develop religious or supernatural beliefs about a generative ghost that are culturally atypical. 
Such beliefs may carry risks such as alienating the mourner from living companions, delayed accommodation, over-reliance on the generative ghost for life advice (i.e., treating it as an oracle), and/or a propensity to unquestioningly carry out actions suggested by the generative ghost (potentially including actions detrimental to themselves or others). 

An additional mental health risk is that of \textit{second deaths} or second losses, the experience of grieving a loved one again.\footnote{Grief scholarship sometimes uses ``secondary loss'' to refer to experiences of grief that follow the ``primary loss'' of a loved one (e.g., loss of relationships, home, shared activities). Here, however, we focus on circumstances that might result in people engaging with their primary loss and grief a second time.} In digital contexts, second deaths occur when data becomes unavailable either through technical obsolescence, deletion, or lack of access  \cite{bassettCreationInheritanceDigital2022}, such as the sudden deprecation of MySpace features that effectively eliminated memorial messages from profiles \cite{brubakerDeathIdentitySocial2015}.
In the case of generative ghosts, second death refers to the emotional harm that may result if a ghost ceases to exist. Second deaths may occur for myriad reasons, including economic (i.e., the company that maintains the generative ghost service goes out of business; survivors' or the deceased's estate can no longer afford maintenance fees),
regulatory (i.e., a government outlaws generative ghosts, which may be particularly likely in theocracies where AI afterlives threaten established belief systems), technological (i.e., obsolete generative ghosts no longer run on future infrastructure), or security reasons (i.e., a hacker disables a generative ghost). While second loss is not unique to AI systems, the potential intensity of engagement that generative ghosts can support means that the loss experienced at the disabling of a generative ghost might be especially profound.

\subsubsection{Reputational Risks}
By \textit{reputational risks} of generative ghosts, we refer to situations in which the generative ghost's interactions might tarnish the memory of the deceased. In addition to altering people's perceptions of the deceased in a manner they would not have desired, reputational risks may also harm the living, either via association with (the now tarnished) deceased (e.g., ``your grandfather was racist'') or through harming their mental health (e.g., a son learning from a generative ghost that his father strongly preferred his sibling over him). We identify three classes of reputational risk: privacy risks, hallucination risks, and fidelity risks.

\textit{Privacy Risks} refer to scenarios where a generative ghost exposes true information that the deceased would not have wanted to be revealed (e.g., because the information is embarrassing, criminal, hurtful to loved ones, etc.). While some privacy risks may be anticipated and prevented via rules when creating a ghost (i.e., ``don't tell my spouse about my affair''), revelations that arise from generated content may be more difficult to prevent (e.g., if the AI correctly infers and reveals the deceased's sexual orientation based on patterns in their data, even though they were closeted while alive). 
Likewise, what constitutes a privacy risk changes over time. 
Aspects of the deceased's life or beliefs that were socially acceptable during their lifetime may come to be frowned upon by future societies (e.g., perhaps in an environmentally conscious future, people who consumed certain kinds of meat or used certain levels of fossil fuels might be retroactively viewed as immoral). 

Privacy risks are context-dependent \cite{nissenbaumPrivacyContextTechnology2009}. For instance, it may be acceptable for a generative ghost to share sensitive information with some audiences, but not with others (e.g., perhaps it is acceptable for the ghost to discuss the deceased's religious beliefs with their spouse, but not with their close work friend). The creation of multiple ghosts, each with different knowledge or abilities targeted toward different audiences, might mitigate privacy risks related to context collapse \cite{davisContextCollapseTheorizing2014,littKnockKnockWho2012a,marwickTweetHonestlyTweet2011}. 

\textit{Hallucination risks} arise when a generative ghost reveals false (but plausible-sounding) information about the deceased. As with privacy risks, such information may tarnish the memory of the deceased (in this case, without true cause), as well as harm the mental well-being of loved ones who believe the falsehoods. Hallucination risks might arise unintentionally (i.e., due to a failure of the underlying AI technology) or might occur due to malicious activity, such as the hacking or hijacking of a generative ghost in order to harm the represented individual or their survivors.

\textit{Fidelity risks} refer to challenges that might arise from both accurate and inaccurate information. Related to hallucination risks, ghosts providing inaccurate information can be a problem, particularly in historical, legal, and economic applications. Yet \textit{accurate} information also comes with risks. As with human memories, personal and family stories change over the years, and most forms of media decay over time. However, digital media defaults towards persistence, which can impede the important roles that forgetting \cite{mayerSchonbergerDeleteVirtueForgetting2009} and evolving memories can play \cite{pasupathiSocialConstructionPersonal2001}. Generative ghosts might disrupt the natural tendency for loved ones to recall the deceased through ``rose-colored glasses'' (i.e., the tendency to focus on positive memories and attributes) by persistently reminding mourners of particular negative events or character traits. 

\subsubsection{Security Risks}

Generative ghosts introduce several \textit{security risks}, including post-mortem identity theft, hijacking, and malicious ghosts. 

Generative ghosts may inspire new twists on crime, such as variants of \textit{identity theft}. Ghosts present more than previously documented security concerns with online data after we die \cite{locasto_security_2011}. Identity thieves may be able to interact with generative ghosts and use innocuous or jailbreak prompts to cause the ghost to reveal sensitive information and even raw data \cite{nasr2023scalable} that might be used for direct financial gain (e.g., passwords, financial information, the location of valuables) or for indirect financial gain (e.g., sensitive personal information about the deceased or the living that might be used in blackmail schemes, phishing attacks, etc.). 

While thieves might only aim to extract useful information,
some criminals may engage in \textit{hijacking} attacks, wherein they take control of a generative ghost. Hijacking attacks might take several forms. One possibility is a ransomware-style attack in which hijackers disable access to a generative ghost until mourners pay them a bounty. Hijackers may also surreptitiously change the functionality of a generative ghost in order to manipulate or harass mourners; this might be accomplished through a variety of methods including modifying source code, prompt injection attacks via the conversational interface, and/or puppetry attacks in which people believe they are chatting with the ghost but are instead chatting with a hijacker posing as the ghost or with a distinct bot the hijacker has substituted for the original generative ghost.  

While identity theft and hijacking represent attacks by nefarious third parties, a first-party security risk is the creation of a \textit{malicious ghost}. Malicious ghosts are first-party generative ghosts whose creators explicitly design them to engage in unpleasant or criminal activities. For example, an abusive spouse might develop a generative ghost that continues to verbally and emotionally abuse their surviving family members. In addition to ghosts that might engage in post-mortem harassment, stalking, trolling, or other forms of abuse of the living, malicious ghosts might be designed to engage in illicit economic activities as a way to earn income for the deceased's estate or to support various causes including potentially criminal ones. 

\subsubsection{Socio-cultural Risks}

Finally, we note that generative ghosts may introduce \textit{socio-cultural risks}. While extremely important, this category of risk is also particularly speculative as it is difficult to anticipate how novel technologies may impact society at scale. For example, the widespread adoption of generative ghosts might cause profound changes to cornerstones of modern society such as the labor market, interpersonal relationships, or religious institutions. While it is difficult to anticipate such societal impacts, even recognizing the possibility of systemic societal change without being confident of its precise direction can help us to be more thoughtful in the design of novel technologies. 

A property of generative ghosts (vs. other types of griefbots that might be limited to simple chat functionalities) is their agentic nature, including the ability to execute actions on behalf of the deceased. For example, while on an individual scale we might view the ability of a generative ghost to earn an income that supports surviving family members (or other causes important to the deceased) as beneficial, at a societal scale this may impact the economy in complex ways that are difficult to anticipate. Widespread economic activity by generative ghosts might impact wages and employment opportunities for the living. Reliance on generative ghosts in some sectors of the economy might result in cultural and economic stagnation if such agents are not capable of developing creative ideas in the same manner as the living or remain anchored to ideas or values from the past.  

Another example of how generative ghosts may impact socio-cultural practices is in how they may alter interpersonal relationships and attendant social structures. For example, in the past two decades, we have seen profound changes in social practices resulting from the widespread adoption of novel technologies such as smartphones and social media.
The age of AI agents is likely to result in similar disruptions of status quo socialization and relationship patterns \cite{openAIagentic}, and generative ghosts may be more disruptive to existing relationship patterns than general-purpose agents because they are modeled on specific loved ones rather than fictional characters or archetypes, possibly making them more compelling and potentially more likely to substitute for traditional relationships. 

Because rituals and beliefs around death are often intertwined with religion \cite{deathReligion}, we anticipate that the widespread adoption of generative ghosts might change religious practices, potentially in profound ways. Possible impacts might include the evolution of practices within a particular religion (i.e., updated rituals), the dissolution of existing religions (i.e., if the existence of AI afterlives alters people's faith in a particular dogma), or the creation of novel religions (i.e., particularly religions that might deify generative ghosts and/or religions that might confer special status upon people with the technical skills to create AI afterlives).
We anticipate that major world religions might issue guidance regarding the use of generative ghosts, and perhaps even offer customized versions of such technologies that are modified to support interaction styles and interface metaphors aligned with particular belief systems.

\section{Discussion}
\label{discussion}




In this paper, we defined the concept of \textit{generative ghosts}, agentic griefbots that may execute a variety of behaviors and actions on behalf of a deceased individual, including a design space of the many design dimensions such technologies may entail. Although digital memorials and simple griefbots have existed for years, generative AI is fundamentally changing the nature and scale of postmortem representations, making generative ghosts a distinct and novel source of AI ethics concerns and a likely cause of social change. As documented in the Related Work section, the recent availability of powerful generative AI models has led to an increase in tech-savvy individuals creating generative-ghost-like systems and startup companies purporting to offer such services; the accelerating interest in post-mortem agents suggests this is an important area for technologists, ethicists, and policymakers to anticipate, monitor, and address. We contributed a design space of key design dimensions of generative ghosts and used this analytic framework to systematically identify potential benefits and risks of this new paradigm of post-mortem representation. Here, we reflect on the importance of careful partnership between the AI and HCI communities to design generative ghosts in a manner that mitigates risk, discuss the need to develop policies that protect values around privacy, consent, and safety, discuss the complexities of foreseeing the societal impacts of ``AI Afterlives,'' and identify key avenues for future research. 

\subsection{Interfaces to Mitigate Risk}
Choices regarding the design dimensions we outline in this paper may influence the benefit/risk landscape of this emerging design space. Careful attention to the interaction design and interfaces of generative ghosts is vital for ensuring that these systems empower representees and the bereaved, and maximize the likelihood of socially beneficial outcomes over risky ones. This includes investing in user studies and social science research to more carefully understand aspects of the design space, such as what interfaces and interactions increase anthropomorphism risks \cite{abercrombie2023mirages, shanahan2023talking, assistantEthics, mimeticModels, aicloneRisks}, and what factors (i.e., features of generative ghosts, contexts of loss, attributes of the bereaved) may contribute to mental health risks from engaging with such technologies. 

Whether a ghost is designed to act as a ``reincarnation'' of or ``represents'' the deceased is a particularly important feature of the design space to consider. Designers will have to make choices about whether a ghost speaks \textit{as} the representee, assuming their voice (which some prior work \cite{brubakerGraveFacebookSite2013} and case studies \cite{romanNews} suggest can be unsettling to some) vs. speaking \textit{about} the representee (ostensibly from the point of view of the system or bot). \textit{Embodiment} can present similar issues, particularly if the ghost's physical design resembles the representee. 

Yet we do not mean to suggest that there are clear right and wrong approaches for each dimension of our design space. After all, a ghost that reincarnates a spouse is likely experienced quite differently than one that reincarnates a historical and/or public figure such as Abraham Lincoln. We suspect that reincarnation-style and embodied interfaces for historical figures carry less risk than for personal friends and family. Characteristics of the representee, the user, and the relationship between the two, will ultimately have profound impacts on how users experience ghosts. However, interaction design can play an important role in framing these experiences.

Given the sensitive nature of generative ghosts, it is particularly important to be vigilant against the use of dark patterns \cite{darkPatterns} in the design of these technologies, to minimize the risk of, for instance, the formation of addictive parasocial relationships that might harm mental health. For instance, one might consider what the equivalent of undesirable ``push notifications'' might be for a generative ghost. Perhaps ghosts should only respond to interactions initiated by the living rather than initiating interactions. 

Generative ghosts might even include interface designs to proactively guard against likely harms. For instance, ghosts might monitor the patterns of interaction with the bereaved and analyze these interactions for patterns of overuse or other types of harm. One could imagine scenarios where a system might initiate interventions such as offering referrals to mental health professionals, reducing its own fidelity, or reducing the hours during which it is available. The types of usage patterns that might prompt concern and the appropriate system responses are important areas for future study.  

Another consideration for interface and interaction design is around \textit{transparency} (i.e., whether and how to clearly signal to end-users that the agent is representing someone who is deceased rather than someone who is still living 
\cite{brubaker_legacy_2016}).
Concerns about transparency are already surfacing in adjacent spaces. For example, in 2024 a Polish Radio station broadcasted an interview with Nobel laureate Wislawa Szymborska without revealing that it was a simulated interview with an AI representation of Szymborska, who had died more than a decade earlier \cite{polandNobel}. In China, a company called Super Brain provides a service in which AI replicas of deceased people can make phone or video calls to family members. According to one report, some customers use this service to avoid having to tell an older relative the distressing news that someone has died \cite{chinaGhosts}. Perhaps AI agents based on real people (e.g., "generative clones") should utilize digital watermarking or fingerprinting capabilities that enable interaction partners to verify whether a representee is alive or deceased. Similar technology might also be used to verify whether an agent is or is not authorized by the representee.  

In addition to the interfaces and interactions for the agents themselves, carefully designed interfaces for creation and ongoing management of generative ghosts are vital for mitigating risk. For example, such interfaces may be part of end-of-life or estate planning processes, and need to clearly convey to representees and their survivors policies around data governance, under what conditions a generative ghost should be terminated (and what happens to the associated data and model), and actions a ghost can and cannot take on behalf of the deceased (i.e., Can a generative ghost continue to perform paid labor on behalf of the deceased in their chosen profession? Can it represent the deceased in legal disputes, such as about estate handling? Can it participate in managing trusts or donor-advised-funds on behalf of the deceased? Can it be consulted regarding end-of-life decisions if the representee is medically incapacitated?). Similarly, interfaces to support data capture and curation are vital both for privacy considerations (i.e., the ability to review and delete sensitive data so that it is not part of a ghost's model) as well as to improve the fidelity of generative ghosts (i.e., by supporting the capture of latent data that exists only in the representee's knowledge rather than in existing media). 

Finally, as with any new media or technology, our experience of generative ghosts may change over time as we gain ``literacy'' in this new medium. For instance, we may find that anthropomorphic ghost designs will be less likely to cause harm over time as end-users become more savvy about understanding this interaction metaphor. 

\subsection{Policies to Mitigate Risk}
Third-party generative ghosts may introduce complex ethical and legal concerns regarding privacy and consent. In addition to potentially violating the desires and privacy of the deceased individual, some classes of third-party ghosts (i.e., those developed by fiscal connections or unconnected entities) may offend or distress surviving family and friends of the deceased. For example, the concern that Hollywood studios might use hypothetical future AI resembling our concept of generative ghosts was among the causes of the 2023 SAG-AFTRA strikes.\footnote{sagaftrastrike.org} 
Policy and governance around third-party ghosts may be prudent, specifically when it comes to who is allowed to create ghosts, who is allowed to be represented, and for what purposes. For example, it might make sense to adopt different policies for private individuals vs. public figures. Distinctions between policies regarding generative ghosts of distant historical figures and public figures whose deaths are more recent may be important as well, particularly since the latter may have living relatives who may be distressed by unauthorized third-party ghosts. For instance, in January 2024 a fan of the late comedian George Carlin (who died in 2008) created an unauthorized comedy special called ``I'm Glad I'm Dead'' using AI technology to mimic Carlin's voice and persona.
Carlin's surviving daughter was highly distressed by this incident \cite{carlinNews},
raising a host of complex ethical and legal issues.  

Of course, status-quo mortality technologies already raise many concerns regarding consent, privacy, and data governance \cite{deathGlitch}. However, these issues are likely to be amplified by the rapid pace at which generative AI technologies are progressing (which makes it difficult for policymakers to keep up with, much less proactively anticipate, risks). These issues are also likely to be exacerbated by the large amount of highly personal data such models will by necessity access in order to create high-fidelity agentic representations of deceased individuals. Policymakers may need to create frameworks that address questions such as under what (if any) circumstances third-party ghosts are permitted (e.g., for public figures), if and how representees or their survivors can terminate a generative ghost (should they change their mind), and what, if any, obligations hosting services have to provide models or data to representees or their survivors in the event of service termination (i.e., due to discontinued products \cite{networkedHeirlooms}, failure of an estate to pay for ongoing services, etc.).  

It may be important 
to have an emergency override -- i.e., a secure and reliable override for temporary or permanently disabling them in the event of malicious third-party events such as hacking or first-party events such as a generative ghost that has been programmed to harass the living. 

Finally, many of these policy considerations point to the importance of considering the lifespans of generative ghosts. Short-term representations might be appropriate for supporting immediate grieving practices or managing an estate and personal affairs. In contrast, longer-term (or indefinite) lifespans are likely important for archival or educational purposes, such as preserving the legacy of a cultural figure for future generations. Policy will necessarily need to vary depending on purpose and lifespan, balancing preferences of the representee, users, societal values, and considerations around data privacy and maintenance responsibilities.



\subsection{Societal Impacts}

It is difficult to anticipate the complex societal changes that may result from technologically and psychologically powerful innovations such as generative ghosts. It is unclear, for instance, what the adoption curve of generative ghosts may be, and what factors might influence this adoption curve; in addition to typical factors influencing technological adoption, such as features and costs, social factors such as network effects are likely important, as well as factors such as changing legal frameworks, religious edicts, and other considerations that are difficult to predict. Generative ghosts exemplify why research must consider more than just the capabilities of novel AI systems. Studying their effects in the context of individual human-AI interactions and their potential systemic, societal-level effects are necessary for robust safety evaluations \cite{weidinger2023sociotechnical}. 

The design choices of early or popular generative ghost services may influence their adoption (or rejection) by segments of society. For instance, third-party generative ghosts might be viewed more negatively than tools that support first-party creation, due to the privacy and consent concerns surrounding the former. 

Other technological trends and developments may also influence likely adoption. For instance, if advances in machine learning improve factuality (i.e., reducing hallucination), thus reducing some reputational risks of generative ghosts, that may encourage adoption. If personal AI agents such as generative clones become commonplace, then transitions from pre-mortem to post-mortem agents may become appealing to a larger audience. If augmented reality or other metaverse technologies see widespread adoption for social and entertainment purposes, then embodied generative ghosts may be more widely accepted.  

If generative ghosts become popular, their associated costs may have significant societal impacts. The economic costs of maintaining an increasing number of high-fidelity generative ghosts may become prohibitive, and the associated computational resources (and associated energy consumption) may also be cause for concern at scale, though such economic, storage, and environmental costs might be mitigated by future advances in 
technology. In addition to considering the cumulative costs at scale, the costs of such services for any given individual may also create new types of digital divides -- it may be prohibitive for families of lower socioeconomic status to afford to create and/or maintain generative ghosts of their loved ones (or they may go into debt to do so if generative ghosts attain important cultural or religious significance). 

\subsection{Future Work}
This conceptual paper takes an analytical approach toward identifying the design space of generative ghosts and the associated benefits and risks of this paradigm. 
As generative AI technologies progress (i.e., improvements in model capabilities, reductions in cost), it will become increasingly feasible to prototype high-quality generative ghosts, which will enable user studies with various stakeholders (e.g., potential representees, their loved ones, clergy, legal experts, etc.) to gain insight into user responses to particular implementations, and to understand patterns of use. In advance of building prototype systems, we recommend investment in research that gathers additional reactions to and requirements for generative ghosts from key stakeholder groups via interviews, surveys, and/or participatory design techniques. Because practices around death and dying are highly culturally dependent, there is a strong need for investigations of generative ghosts to engage with diverse user populations beyond the traditional ``WEIRD'' (Western, Educated, Industrialized, Rich, and Democratic) demographic. Including samples that vary along dimensions such as age, religion, and nationality is particularly important. 

This paper identifies an emerging trend (creation of generative ghosts) that is likely to scale quickly due to (1) the fast pace of advances and availability of powerful generative AI models that will empower tech-savvy individuals to create generative ghosts and (2) the emergence of start-ups that will provide ghost-creation services to a broader consumer market. In addition to describing an emerging phenomenon, this paper is a call-to-action to scholars to collect empirical data that can be used to understand more deeply how choices in the design of generative ghosts (i.e., our framework of design dimensions) might impact outcomes (i.e., our risk/benefit analysis). 
The benefits and risks of generative ghosts will likely arise from the intersection of these design dimensions and contextual factors (such as personal characteristics of the bereaved, length of time since the deceased has passed, and culture).
As such, contextual factors are also an important aspect of future experimental designs. The findings of such studies can then be used to provide design guidelines that can steer such technologies in beneficial directions, as well as provide information that can support policymakers and other stakeholders in understanding what societal frameworks may be needed to adapt to this phenomenon.

\section{Conclusion}
\label{conclusion}

We introduced the concept of \textit{generative ghosts} --- agentic, AI-powered representations of deceased individuals. Using an analytical approach, we proposed a set of design dimensions of generative ghosts (provenance, deployment timeline, anthropomorphism paradigm, multiplicity, cutoff date, embodiment, and representee type), and reflected on how combinations of these choices might introduce both benefits and risks for the representee, their loved ones, and society at large. Our paper highlights a growing trend of using AI systems to represent the deceased; our design space can help scholars and policymakers understand how this trend may evolve, and our analysis of potential benefits and risks provides an agenda for researchers and policymakers to conduct studies and collect data that will offer empirical evidence to understand how these design dimensions (as well as other contextual considerations) might relate to risk/benefit trade-offs. This paper sets the stage for a future research agenda bringing together interdisciplinary experts in AI ethics, ML models, Human-AI Interaction, Policy, and other cultural and religious institutions to study and design AI afterlives that balance the benefits and risks of this emerging sociotechnical paradigm.




\bibliographystyle{ACM-Reference-Format}
\bibliography{references}

\end{document}